\documentclass[apj]{emulateapj}
\usepackage{graphicx,natbib,amsfonts,amssymb,rotating,lscape}
\usepackage{apjfonts}
\usepackage{color}
\usepackage{lscape}                                        
\usepackage{graphicx}
\usepackage{natbib}
\newcommand{\kms}{\ifmmode {\rm km\ s}^{-1} \else km s$^{-1}$\ \fi}
\newcommand{\ergs}{\ifmmode {\rm erg\ s}^{-1} \else erg s$^{-1}$\ \fi}
\newcommand{\msun}{\ifmmode M_{\odot} \else $M_{\odot}$\ \fi}

\newcommand{\feii}{Fe {\sc ii}\ }
\newcommand{\mgii}{Mg {\sc ii}\ }
\newcommand{\civ}{C {\sc iv}\ }

\newcommand{\lb}{\ifmmode L_{\rm Bol} \else $L_{\rm Bol}$\ \fi}
\newcommand{\ledd}{\ifmmode L_{\rm Edd} \else $L_{\rm Edd}$\ \fi}
\newcommand{\lx}{\ifmmode L_{\rm 2-10keV} \else  $L_{\rm 2-10keV}$\ \fi}
\newcommand{\hb}{\ifmmode H\beta \else H$\beta$\ \fi}
\newcommand{\hbb}{\ifmmode H\beta^{b} \else H$\beta^{b}$\ \fi}
\newcommand{\hbn}{\ifmmode H\beta^{n} \else H$\beta^{n}$\ \fi}
\newcommand{\ha}{\ifmmode H\alpha \else H$\alpha$\ \fi}
\newcommand{\oiii}{[O {\sc iii}]\ }

\newcommand{\mbh}{\ifmmode M_{\rm BH}  \else $M_{\rm BH}$\ \fi}
\newcommand{\lv}{\ifmmode f_{\lambda}(5100\AA) \else $f_{\lambda}(5100\AA)$\ \fi}
\newcommand{\llv}{\ifmmode \lambda L_{\lambda}(5100\AA) \else $\lambda L_{\lambda}(5100\AA)$\ \fi}

\newcommand{\mdot}{\ifmmode \dot{m} \else \dot{m} \fi }
\newcommand{\llog}{\ifmmode {\rm log} \else {\rm log} \fi }
\newcommand{\lbol}{L_{\rm bol}}
\newcommand{\hei}{He {\sc i}\ }
\newcommand{\heii}{He {\sc ii}\ }
\newcommand{\hg}{\ifmmode H\gamma \else H$\gamma$\ \fi}
\renewcommand{\baselinestretch}{1.15}

\begin{document}
\title{The variability of optical \feii emission in PG QSO 1700+518}
\author{Wei-Hao Bian$^{1}$, Kai Huang$^{1}$, Chen Hu$^{2}$, Li Zhang$^{1}$,
Qi-Rong Yuan$^{1}$, Ke-Liang Huang$^{1}$, and Jian-Min Wang$^{2}$\\
$^{1}$Department of Physics and Institute of Theoretical Physics,
Nanjing Normal University, Nanjing
210097, China\\
$^{2}$Key Laboratory for Particle Astrophysics, Institute of High
Energy Physics, Chinese Academy of Sciences, Beijing 100039,
China\\
} \shorttitle{\feii variability in PG QSO 1700+518}
\shortauthors{Bian, et al.}
\begin{abstract}
It is found that \feii emission contributes significantly to the
optical and ultraviolet spectra of most active galactic nuclei. The
origin of the optical/UV \feii emission is still a question open to
debate. The variability of \feii would give clues to this origin.
Using 7.5 yr spectroscopic monitoring data of one Palomer-Green (PG)
quasi-stellar object (QSO), PG 1700+518, with strong optical \feii
emission, we obtain the light curves of the continuum \lv, \feii,
the broad component of \hb, and the narrow component of \hb by the
spectral decomposition. Through the interpolation cross-correlation
method, we calculate the time lags for light curves of \feii, the
total \hb, the broad component of \hb, and the narrow component of
\hb with respect to the continuum light curve. We find that the
\feii time lag in PG1700+518 is $209^{+100}_{-147}$ days, and the
\hb time lag cannot be determined. Assuming that \feii and \hb
emission regions follow the virial relation between the time lag and
the FWHM for the \hb and \feii emission lines, we can derive that
the \hb time lag is $148^{+72}_{-104}$ days. The \hb time lag
calculated from the empirical luminosity--size relation is 222 days,
which is consistent with our measured \feii time lag. Considering
the optical \feii contribution, PG 1700+518 shares the same
characteristic on the spectral slope variability as other 15 PG QSOs
in our previous work, i.e., harder spectrum during brighter phase.
\end{abstract}


\section{Introduction}
The variability is a common phenomenon in quasi-stellar objects
(QSOs) and provides a powerful constraint on their central engines.
In the past two decades, the optical variability research focused on
the spectral monitoring instead of the pure photometric monitoring.
With the active galactic nuclei (AGNs) watch and the Palomer-Green
(PG) QSOs spectrophotometrical monitoring projects, the
reverberation mapping method, i.e., exploring the correlation
between the emission lines and the continuum variations, is used to
investigate the inner structure in AGNs \citep[e.g.,][]{Blandford82,
Peterson93}. It is found that motions of clouds in the broad line
regions (BLRs) are virialized \citep[e.g.,][]{Kaspi00, Kaspi05,
Peterson04}.  With the line width of \hb, \mgii, \civ from BLRs, the
empirical size-luminosity relation derived from the mapping method
is used to calculate the masses of their central supermassive black
holes \citep[SMBHs; e.g.,][]{Kaspi00, McLure04, Bian04, Peterson04,
Greene05}.

It is found that the \feii emission contributes significantly to the
optical and ultraviolet spectra of most AGNs. Thousands of UV \feii
emission lines blend together to form a pseudocontinuum, resulting
in the ``small blue bump'' around 3000\AA\ when they are combined
with Balmer continuum emission \citep[e.g.,][]{Wills85}. The optical
\feii would lead to two bumps in two sides around the \hb $\lambda
4861$\AA\ \citep[e.g.,][]{Boroson92}. It is found that the flux
ratio of \feii to \hb, $R_{\rm Fe}$, where the optical \feii flux is
the flux of the \feii\ emission between $\lambda$4434 and
$\lambda$4684, strongly correlates with the so-called Eigenvector 1,
which is suggested to be driven by the accretion rate
\citep[e.g.,][]{Boroson92, Marziani03a}.

The origin of the optical/UV \feii emission is still an open
question. It is found that photoionized BLRs cannot produce the
observed shape and strength of the optical \feii emission and that
the strength of UV \feii cannot be explained unless considering the
micro-turbulence of hundreds of \kms or the collisional excitation
in warm, dense gas \citep{Baldwin04}. However, \citet{Vestergaard05}
found the correlation between the optical \feii variance and the
continuum variance and suggested that the optical \feii is due to
the line fluorescent in a photoionized plasma. It suggests that the
optical \feii line do not come from the same region as the UV \feii
emission \citep[e.g.,][]{Kuehn08}. \citet{Maoz93} found that the
reverberation time lag of UV \feii in NGC 5548 is about 10 days,
similar to \civ time lag, smaller than the \hb time lag. The
reverberation measurement for the optical \feii emission has not
fared so well. Some suggested that the optical emission is produced
in the same region as the other broad emission lines, and some
suggested that it is in the outer portion of the BLRs because of
narrower FWHM of \feii with respect to \hb \citep[e.g.,][]{Laor97,
Marziani03a, Vestergaard05, Kuehn08}. Recently, \citet{Hu2008a,
Hu2008b} did a systematic analysis of \feii emission in QSOs from
the Sloan Digital Sky Survey (SDSS), and found that the \feii
emission is redshifted with respect to the rest frame defined by the
\oiii narrow emission line and \hb intermediate-width component is
correlated with \feii which locates at the outer portion of the
BLRs.

\citet{Kaspi00} gave the 7.5 yr spectroscopic monitoring data for 17
PG QSOs. There is one PG QSO, PG 1700+518, with strongest optical
\feii emission and $R_{\rm Fe}=1.42$ \citep{Turnshek85, Boroson92}.
Its \hb FWHM is $1846\pm 682 \kms$ \citep{Peterson04}, and it is
also called as a narrow line Seyfert 1 galaxy (NLS1). Using the
\feii template from one NLS1, I ZW 1, we model the \feii emission to
investigate the \feii variability and the relation to the continuum
variability in PG 1700+518. The data and analysis are described in
Section 2, the results are given in Section 3, the discussion is
given in Section 4, and the conclusions are presented in Section 5.
All of the cosmological calculations in this paper assume $H_{\rm
0}=70 \kms \rm ~Mpc^{-1}$, $\Omega_{\rm M}=0.3$, $\Omega_{\Lambda} =
0.7$.

\section{The data and analysis}
   \begin{figure}
   \centering
   \includegraphics[height=8cm,width=6cm,angle=-90]{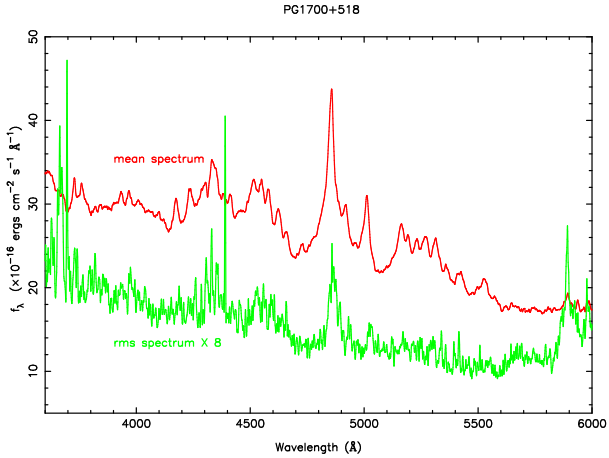}
   \caption{Mean spectrum (the red curve) and rms
   spectrum (the green curve; multiplied by 8) for PG 1700+518.}
   \end{figure}

The spectroscopic monitoring data of PG 1700+518 cover 7.5 years
from 1991 to 1998, which were done every 1--4 months by using 2.3 m
telescope at the Steward Observatory and 1 m telescope at the Wise
Observatory. The total number of optical spectra for PG 1700+518 is
39. The observational wavelength coverage is from $\sim$ 4000 to
$\sim$ 8000 \AA\ with a spectral resolution of $\sim$ 10\AA. Spectra
were calibrated to an absolute flux scale using simultaneous
observations of nearby standard stars \citep{Kaspi00}. The 39
spectra of PG 1700+518 are available on the Web site {\footnote{
http://wise-obs.tau.ac.il/$\sim$shai/PG/}}.

In order to check its spectral variance, we calculate its mean and
rms spectra \citep{Kaspi00, Peterson04}. In Figure 1, the mean
spectrum (the red curve) shows strong optical \feii emission, and
the rms spectrum (the green curve) shows variable emission for \hb
$\lambda 4861$, \hei $\lambda 5878$, and \feii (the blueward and
redward of \hb $\lambda 4861$). In the rms spectrum, we can find
\feii features at $\sim$ 4500\AA, 4924\AA\ and 5018\AA, suggesting
variable \feii. The part of \feii emission between 4430\AA\ and
4770\AA\ in the rms spectrum is obvious than that between 5080 \AA\
and 5550\AA. It is due to the variable continuum slope, more
variable in blueward of the spectrum. It is consistent with harder
spectrum during bright phase \citep[Figure 6]{Pu06}. In the rms
spectrum, we also find weak \heii $\lambda 4686$.

   \begin{figure}
   \centering
   \includegraphics[height=8cm,width=6cm,angle=-90]{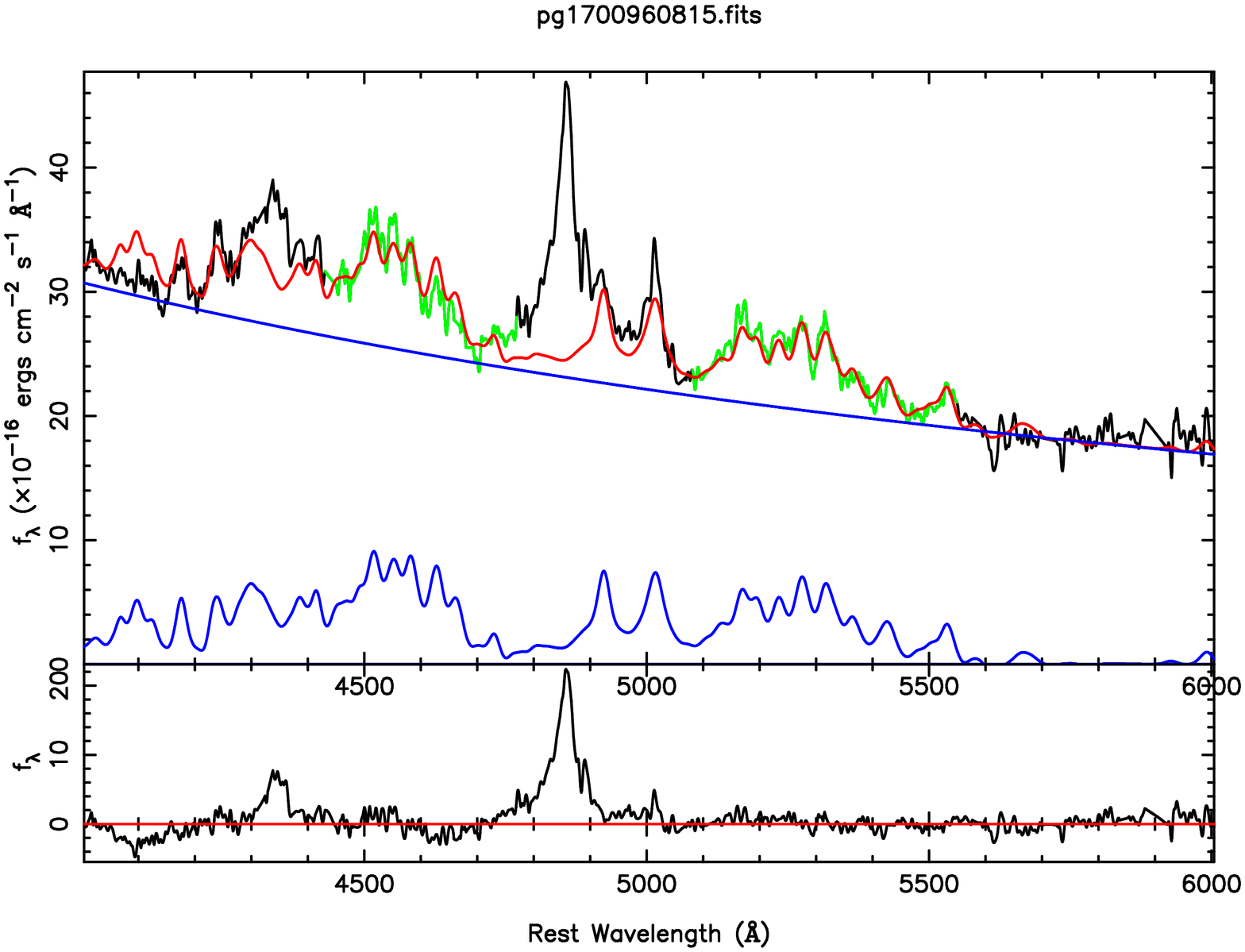}
   \includegraphics[height=8cm,angle=-90]{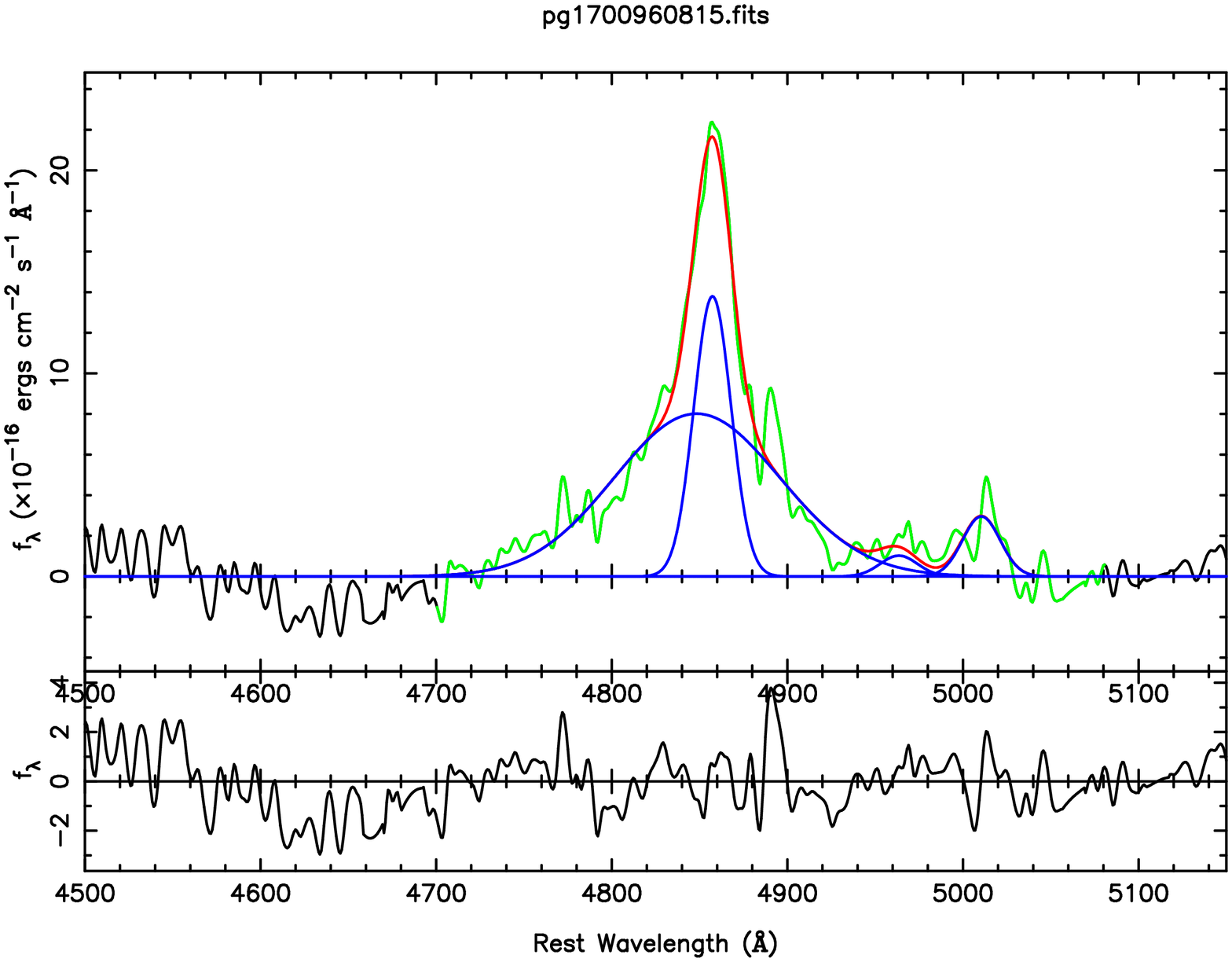}
   \caption{Example spectral decomposition for PG 1700+518.
   In the top panel, the black curve is the
observed spectrum after the corrections of Galactic-extinction and
the redshift, the red line is the sum of the power-law continuum and
\feii multiples (blue curves). The green ranges are our fitting
windows. The bottom panel is the multi-Gaussian fits for the \hb and
\oiii lines. The red line is the sum of all multi-Gaussian (blue
curves). The green curve is our fitting range of the pure \hb and
\oiii emissions after the subtraction of the power-law continuum and
\feii multiples.}
   \end{figure}

We use following steps to do the spectral decomposition, which have
been used to analyze the spectra for a large QSOs sample from SDSS
\citep{Hu2008a, Bian08}.

(1) First, the observed spectra are corrected for the Galactic
extinction using $A_V=0.116$ from the NASA/IPAC Extragalactic
Database (NED), assuming an extinction curve of Cardelli, Clayton \&
Mathis (1989; IR band) and O'Donnell (1994; optical band) with
$R_V=3.1$. Then the spectra are transformed into the rest frame by
the redshift of $0.292$.

(2) The optical and ultraviolet \feii template from the prototype
NLS1 I ZW 1 is used to subtract the \feii emission from the spectra
\citep{Boroson92, Vestergaard01}. The I ZW 1 template is broadened
by convolving with a Gaussian of various linewidths, the centroid
wavelength shifts and fluxes. A power-law continuum is added in the
fitting. The best modeling of the \feii\ and the power-law continuum
is found when $\chi^2$ is minimized in the fitting windows:
4430-4770, 5080-5550\AA\ (see an example fit in Figure 2). The
monochromatic flux at 5100\AA, \lv, is calculate from the power-law
continuum. Because of the spectral coverage, we did not consider
Balmer continuum (see the mean spectrum in Figure 1).

(3) Considering weak \oiii$\lambda \lambda4959, 5007$ lines, two
sets of one Gaussian are used to model them. We take the same line
width for each component, and fix the flux ratio of
\oiii$\lambda$4959 to \oiii $\lambda$5007 to be 1:3. For the
asymmetric profile of the \hb profile, two-Gaussian is used to model
the \hb line,  \hbb and \hbn. The \hbb and \hbn fluxes are
calculated from integrating the corresponding fitting components.
The flux for total \hb, $\rm \hb^{n+b}$,  is the sum of \hbb and
\hbn fluxes.

\section{Result}
\subsection{The light curves for the continuum, \feii, \hb}

   \begin{figure}
   \includegraphics[height=8cm,angle=0]{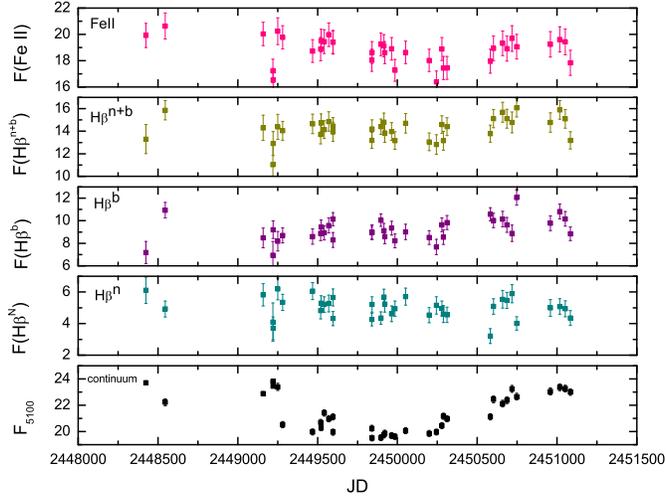}
   \caption{Light curves of \feii, H$\beta^{n+b}$, \hbb, \hbn, \lv (from top to bottom).
   Emission-line fluxes are displayed in units of $10^{-14} \ergs \rm cm^{-2}$, the continuum flux determined
   from the power-law at 5100\AA\ is given in units of $10^{-16} \ergs \rm cm^{-2}$ \AA$^{-1}$.}
   \end{figure}

By IRAF-splot, the signal-to-noise ratios (S/Ns) between 7400\AA\
and 7600\AA\ in the observational frame are measured for these 39
spectra. Two spectra (2nd and 35th spectra: pg1700910712.fits,
pg1700980414.fits) are ignored in our next analysis for their lower
S/N less than 10 \citep{Kaspi00}. For the left 37 spectra, the
distribution of S/N is $25.3\pm 9$. The goodness of modeling of the
\feii and continuum is tested by the elimination of \feii features
at $\lambda 4924$ and $\lambda 5017$ (see Figure 2). We calculate
the \feii flux by integrating the \feii template fit between
$\lambda$4434 and $\lambda$4684. In Figure 3, we show the light
curves of \feii, H$\beta^{n+b}$, \hbb, \hbn, and \lv (from top to
bottom).

We use the normalized variability measure defined by \citet{Kaspi00}
to compare the line variability to the continuum variability,
$\sigma_N=100(\sigma^2-\delta^2)^{1/2}/\bar{f}$, where $\bar{f}$ and
$\sigma$ are the average and the rms of the flux in a given light
curve, and $\delta$ is the mean uncertainty in a given light curve.
The $\sigma_N$ are 6.73, 2.58, 8.46, 8.38, and 3.65 for the light
curves of \lv,  \feii, \hbb, \hbn, and H$\beta^{n+b}$, respectively,
in Figure 3. \citet{Kaspi00} showed that $\sigma_N$ for the light
curves of \lv and \hb are 6.8, 3.2, with respectively, which are
consistent with our results. And the $\sigma_N$ of \feii is smaller
than that for \lv, \hbb, \hbn, and H$\beta^{n+b}$.

\subsection{Time lag from CCCD}
   \begin{figure}
   \includegraphics[height=4cm,width=4cm,angle=0]{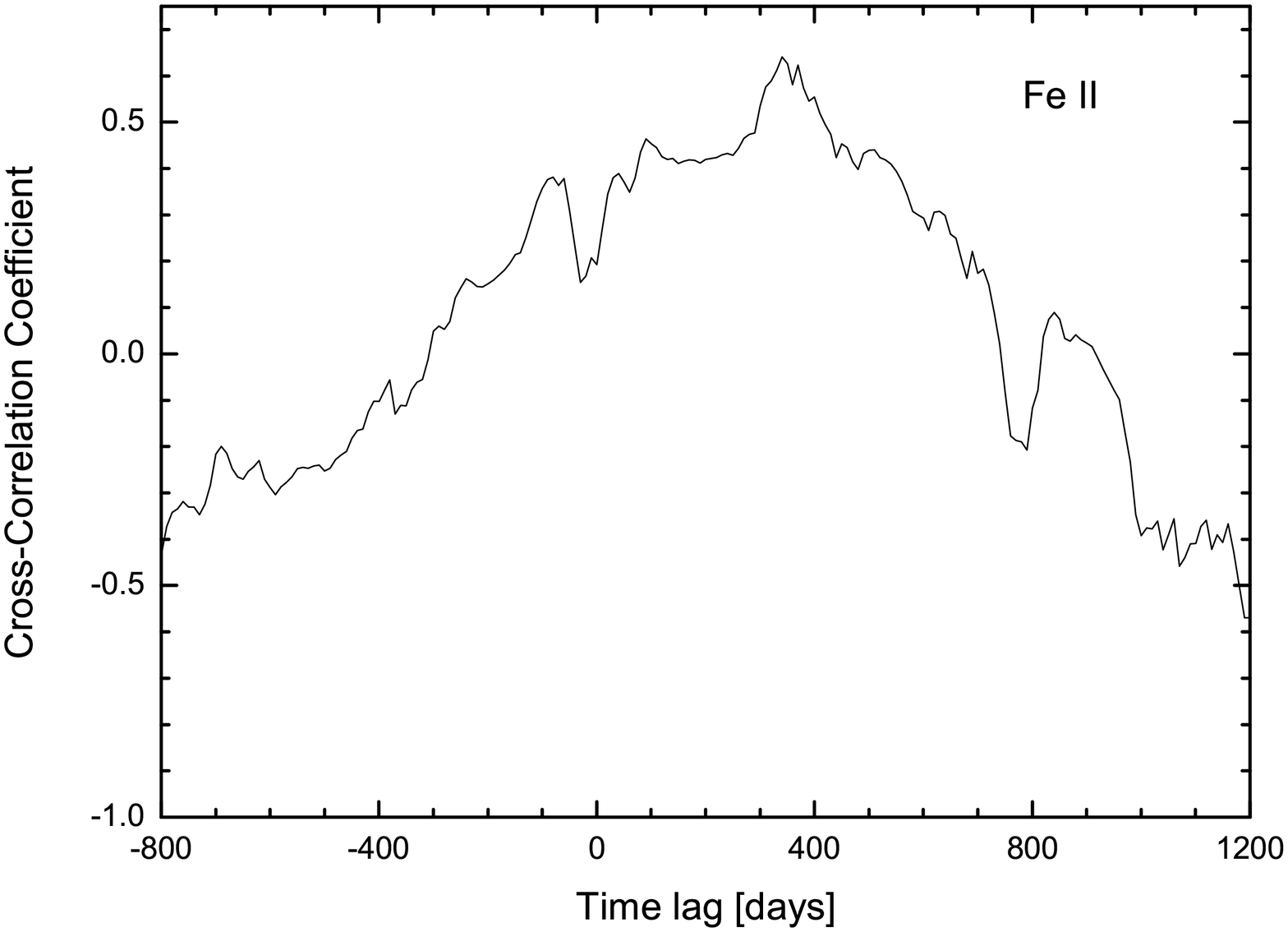}
   \includegraphics[height=4cm,width=4cm,angle=0]{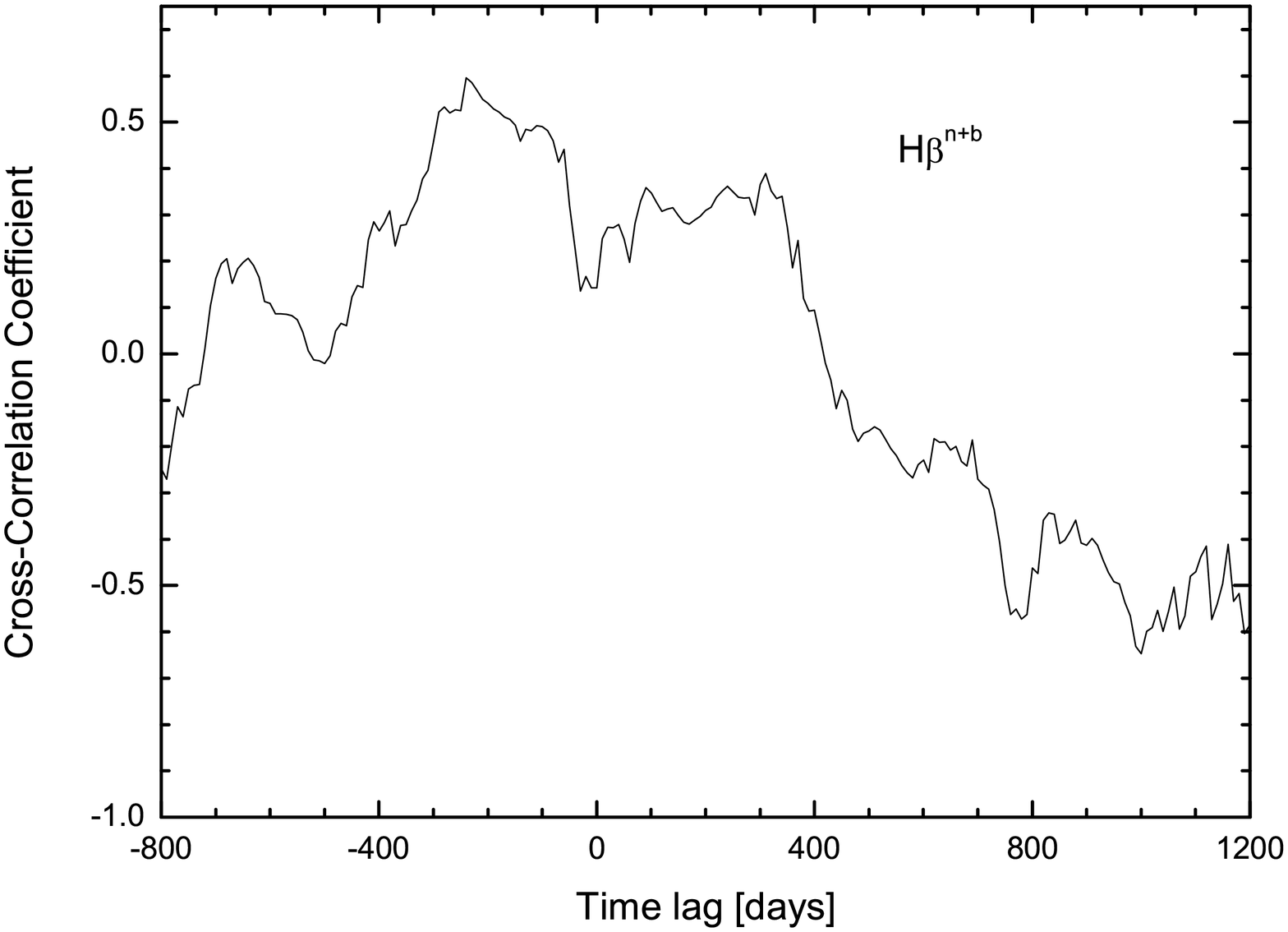}

   \includegraphics[height=4cm,width=4cm,angle=0]{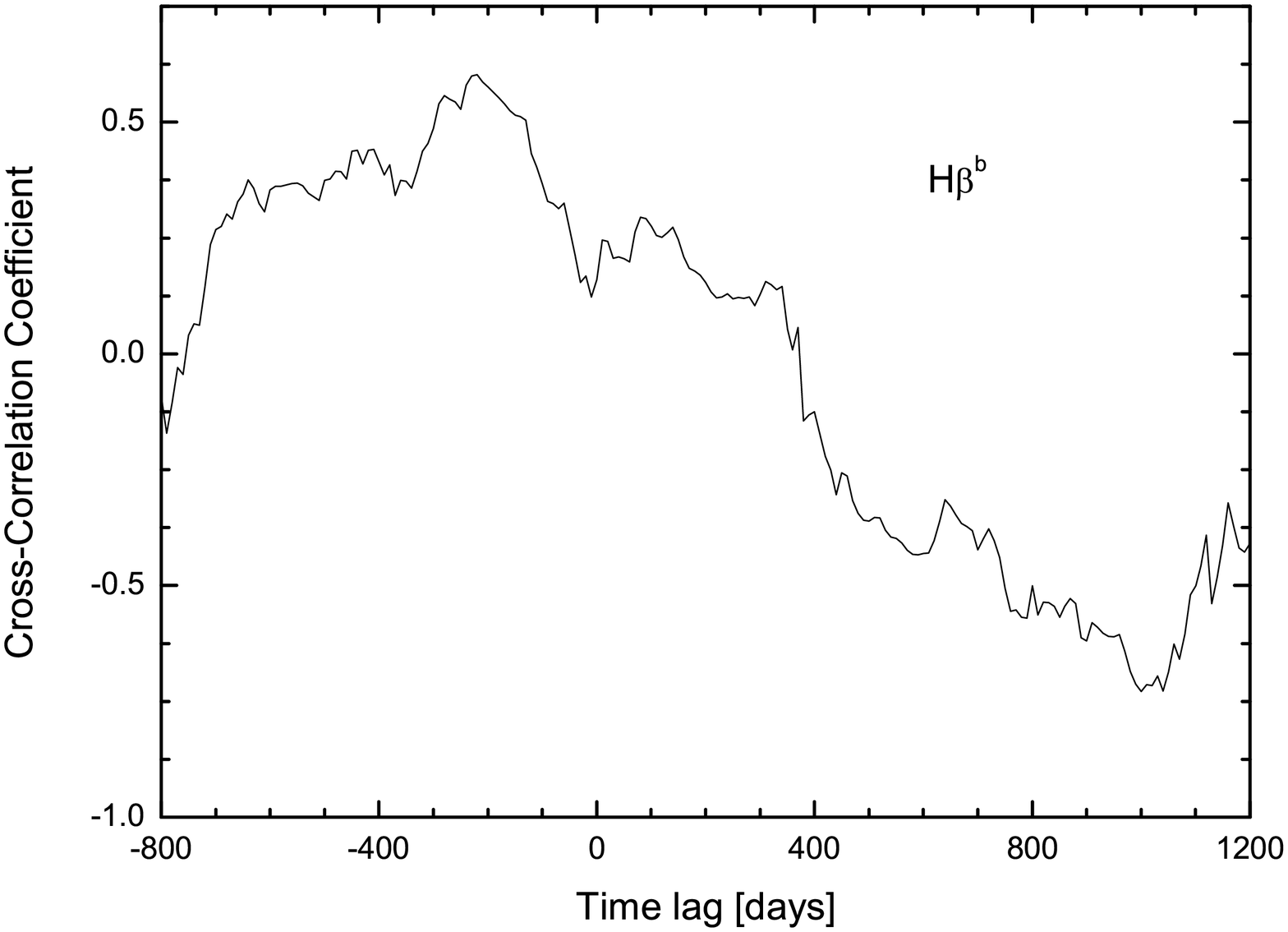}
   \includegraphics[height=4cm,width=4cm,angle=0]{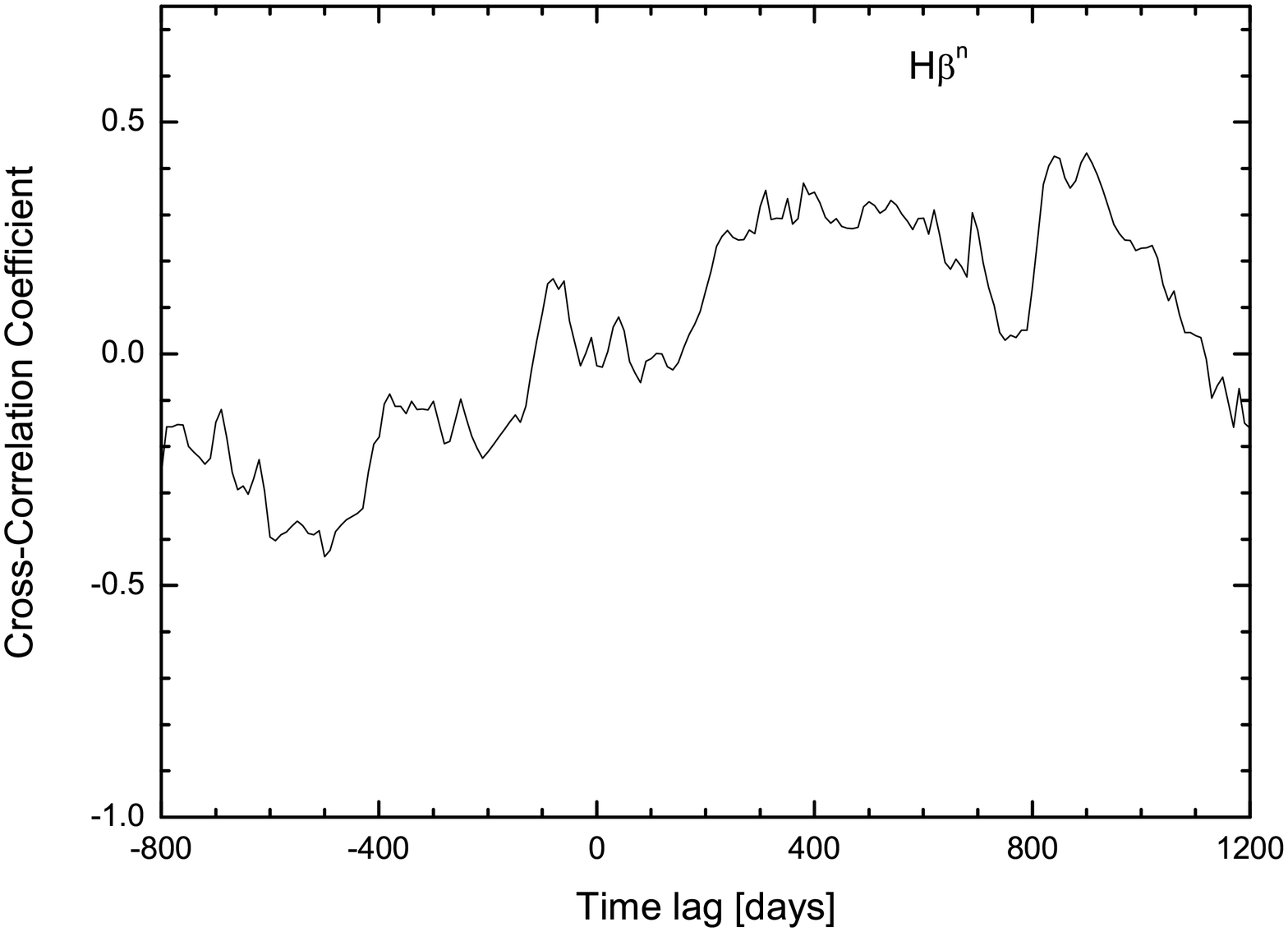}
   \caption{Cross-correlation functions (CCFs) for the
continuum--\feii (top left), the continuum--H$\beta^{n+b}$ (top
right), the continuum--H$\beta^{b}$(bottom left), and the
continuum--H$\beta^{n}$ (bottom right) for PG 1700+518.}
   \end{figure}

   \begin{figure}
   \includegraphics[height=4cm,width=4cm,angle=0]{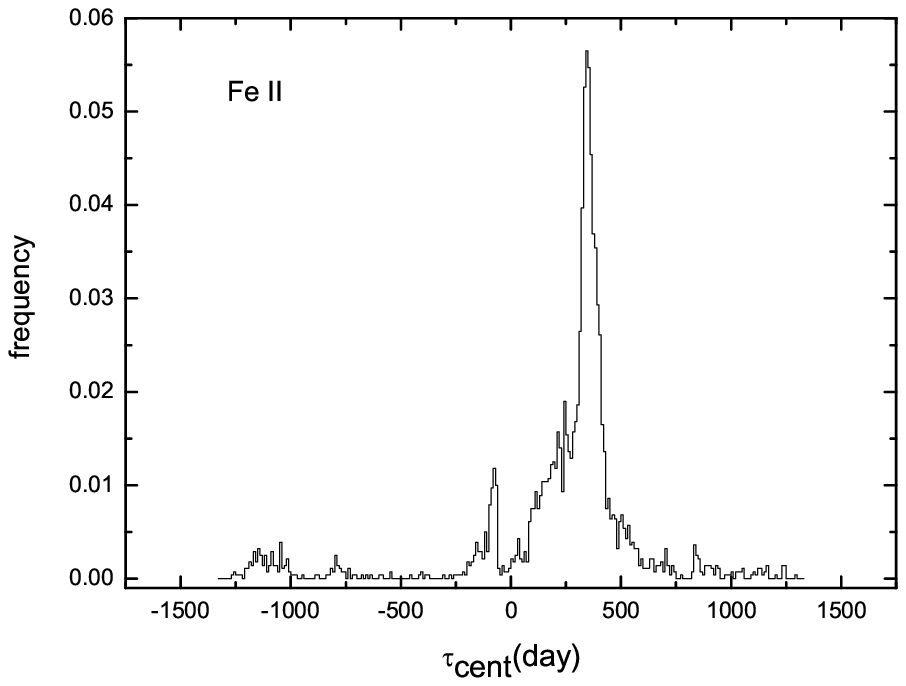}
   \includegraphics[height=4cm,width=4cm,angle=0]{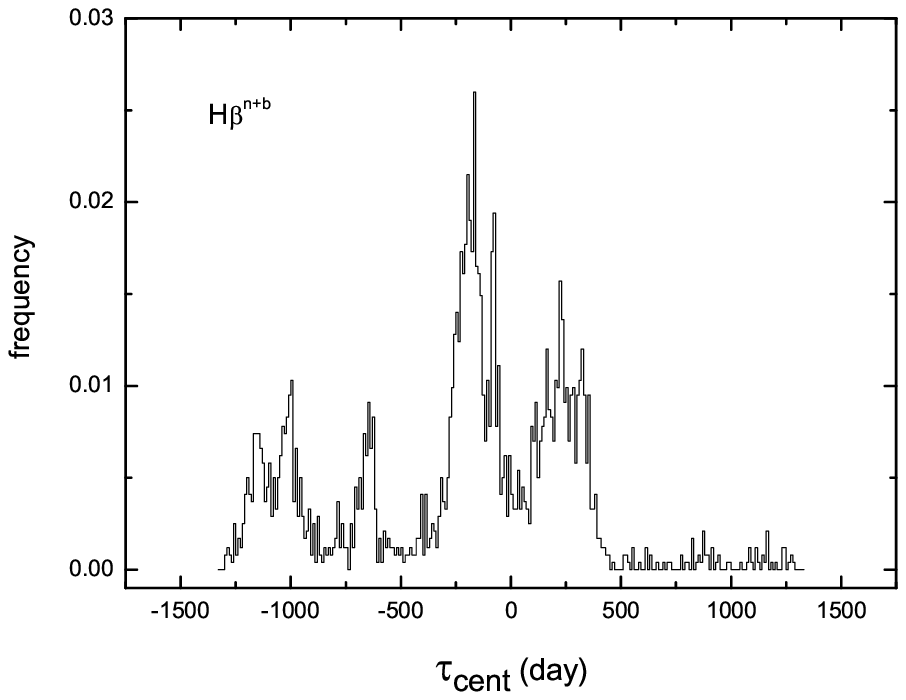}

   \includegraphics[height=4cm,width=4cm,angle=0]{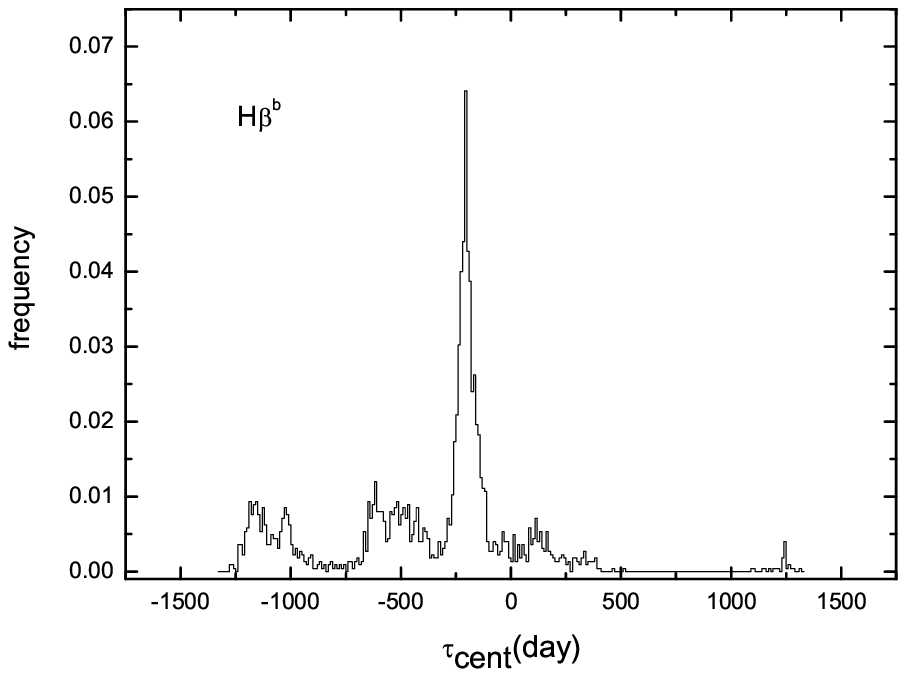}
   \includegraphics[height=4cm,width=4cm,angle=0]{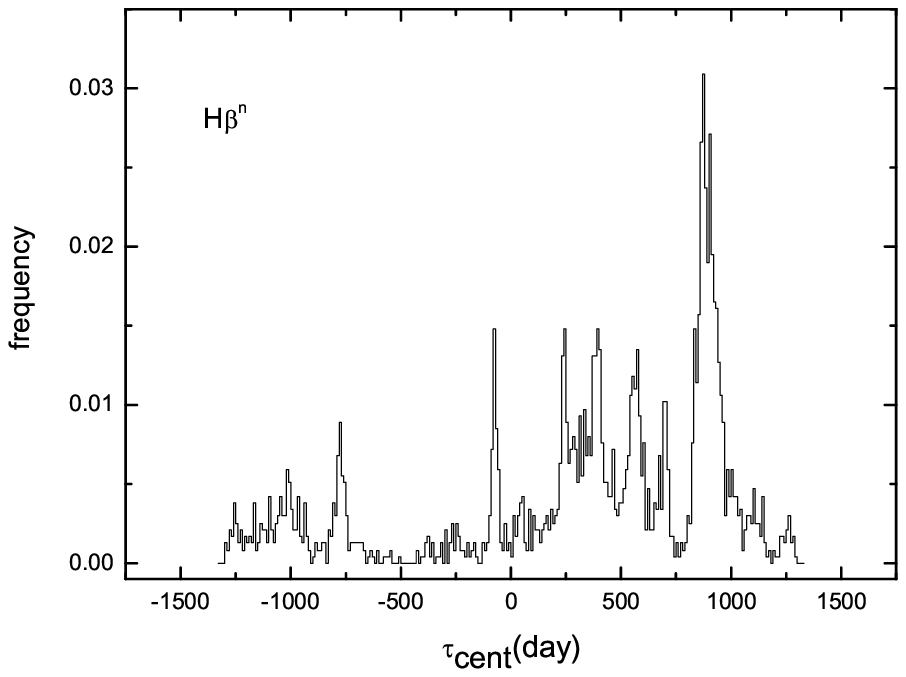}
   \caption{Cross-correlation centroid distributions (CCCDs) for the continuum--\feii cross-correlation (top left),
   the continuum--H$\beta^{n+b}$ cross-correlation (top right), the continuum--H$\beta^{b}$ cross-correlation (bottom
   left), and the continuum--H$\beta^{n}$ cross-correlation (bottom right) for
   PG 1700+518.}
   \end{figure}

The measurement of the line time lag, $\tau$, is made by
cross-correlating the emission line and continuum light curves. In
order to compare with the result of \citep{Peterson04}, we use their
code to do the line lag measurements. In Figure 4, we give the
interpolated cross-correlation functions (CCF) for the
continuum--\feii (top left), the continuum--H$\beta^{n+b}$ (top
right), the continuum--H$\beta^{b}$(bottom left), and the
continuum-H$\beta^{n}$ (bottom right) for PG 1700+518. We find a
peak in the CCF for \feii light curve, which can be used to
determine the time lag for \feii curve (Figures 4, 5). Because of
many peaks with positive lag times and/or peaks with negative lag
times in the CCFs for other three curves, we can't determine their
time lags (Figures 4 and 5).

Through the Monte Carlo FR/RSS method, the uncertainty of $\tau$ can
be determined (Peterson et al. 2004). In the code, we adopt a
minimum correlation coefficient of 0.4, the centroid threshold of
0.8 $R_{\rm max}$, and the number of trials is 3000 (Peterson et al.
2004). In Figure 5, we give the cross-correlation centroid
distributions (CCCDs) for the continuum--\feii cross-correlation
(top left) and the continuum--H$\beta^{n+b}$ cross-correlation (top
right) for PG 1700+518, as well as that for \hbb and \hbn (bottom
panels). It is obvious that the CCCD for \feii have a narrow
positive peak. Following the suggestion given by \citep{Peterson04},
we calculate the mean of the CCCD in all valid trials as the \feii
centroid time lag, as well as its upper and lower uncertainties. We
find that \feii time lag ,$\rm \tau_{\rm Fe II}$, in PG 1700+518 is
$270^{+130}_{-190}$ days. Its mean CCF $R_{\rm max}$ is $0.54\pm
0.08$. There are many positive peaks and/or strong negative peaks
for \hb, \hbn, and \hbb. Therefore, we cannot give the line lags for
\hb, \hbn, and \hbb. With the redshift of 0.292 for PG 1700+815, in
the rest frame, the \feii time lag in PG1700+518 is
$209^{+100}_{-147}$ days. Because the \hb time lag cannot be
determined, we do not know whether the region emitting \feii is
located outside of the region emitting the broad \hb lines
\citep{Marziani03b, Vestergaard05, Popovic07, Hu2008b, Kuehn08}.

\section{Discussion}
\subsection{The \feii fitting method}
In the analysis of the optical \feii light curve for PG 1700+518, we
fit the optical spectrum by the \feii template instead of directly
calculating the optical \feii flux in the selected wavelength range
\citep[e.g.,][]{Kuehn08}. In modeling \feii emission, we
simultaneously model the power-law continuum, which is different
from \citet{Wang05} in the \feii analysis of an NLS1 NGC 4051. We
also consider various FWHMs, centroid wavelength shifts and fluxes
in the convolving of the \feii template. The accuracy of the
measurement for the continuum shape depends on the wavelength
coverage. For these 37 spectra of PG 1700+518, the wavelength
coverage is mainly between $\sim$ 3500\AA and $\sim$ 6000\AA\ in the
rest frame. \citet{Vestergaard05} found that the optical \feii
feature to the blue of \hb is contaminated by \hei $\lambda 4471$
and \heii $\lambda 4686$ lines. For PG 1700+518, the \hei $\lambda
4471$ and \heii $\lambda 4686$ lines are not strong (Figure 1).
Therefore, we use the fitting windows of 4430-4770 and 5080-5550\AA\
to exclude the emission lines of \hb, \hg$\lambda 4340$, and \oiii
$\lambda \lambda$4959, 5007 (Figure 2). When we mask the region of
\heii $\lambda 4686$ in the fitting, the fitting result is almost
the same. We tried \feii template of \citet{Veron-Cetty04}, it is
almost the same to that of I ZW 1.

\subsection{The optical \feii emitting region}
\citet{Kuehn08} presented the reverberation analysis of optical
\feii for Ark 120. They gave the light curves of the blue/red side
of \feii, \hb, and continuum by setting the measurement windows
(Figure 1 in Kuehn et al. 2008). Although the \feii
cross-correlation function is very broad and flat-topped, they
suggested that the optical \feii--emitting region, $\sim 320$ days,
is several times larger than the \hb zone ($\sim 57$
days).\citet{Kuehn08} found that it is difficult to constrain the
FWHM of optical \feii for Ark 120 because of its very smooth \feii
emission (Figure 8 in Kuehn et al. 2008). Modeling the \feii
emission in PG 1700+518, we find that the mean value of \feii FWHM
is $1554\pm 110$ \kms. Because the change of \hb profile due to the
\feii contribution is not too much (Figure 1), we adopted the \hb
FWHM value of $1846\pm 682$ \kms by \citet{Peterson04}. We find that
$\rm (FWHM_{H\beta}/FWHM_{Fe II})^2=1.41$. Assuming that \feii and
\hb emission regions follow the virial relation between the time lag
and the FWHM for the \hb and \feii emission lines, we can derive
that the \hb time lag is $148^{+72}_{-104}$ days. We also find taht
the new estimated \hb time lag is consistent with $R_{\rm BLR} -
\llv$ relation by \citet[see \ their \ Figure \ 5]{Bentz09}.

Considering the host contribution in \lv, \citet{Bentz09} suggested
a new relation between BLRs size and \llv, $\log R_{\rm
BLR}=(-21.3^{+2.9}_{-2.8})+(0.519^{+0.063}_{-0.066}) \log\llv ~~\rm
(lt-days)$. \citet{Kaspi00} gave the average flux of \lv (between
6520 and 6570 \AA) without excluding \feii contribution, $(22.0\pm
1.5)\times 10^{-16} \ergs \rm cm^{-2}$\AA$^{-1}$. After excluding
\feii contribution, we find that the value of \lv at 5100\AA\ is
$(21.4\pm 1.5)\times 10^{-16} \ergs \rm cm^{-2}$\AA$^{-1}$.
Therefore, the \feii correction is very small for \lv. Corrected the
contribution from starlight, \citet{Bentz09} gave \lv as $(18.5\pm
1.5)\time 10^{-16} \ergs \rm cm^{-2}$\AA$^{-1}$ and \llv as
$3.63\times 10^{45} \ergs$ (the starlight contribution is about 16\%
in its total flux). The expected $R_{\rm BLR}$ from $R_{\rm BLR} -
\llv$ relation \citep{Bentz09} and \llv of $3.63\times 10^{45}
\ergs$ is 222 lt-days. Considering the larger uncertainty of
intercept in this relation (about 3 in $\log R_{\rm BLR}$), this
result is consistent with our estimated \hb time lag,
$148^{+72}_{-104}$ days \citep[see \ their \ Figure \ 5]{Bentz09}.

If we take the FWHM/time lag uncertainties into consideration, the
\feii emission region is located near the \hb emission region, not
conclusively located outside of the \hb emission region. Kuehn et
al. (2008) suggested that optical \feii emission is possibly
produced at the dust sublimation radius, $R_{\rm dust}=476 \times
[\lbol/10^{45} \ergs]^{0.5}  ~~\rm (lt-days)$ \citep{Elitzur06}. By
$\lbol=9\llv$, for PG1700+518, we find that the dust sublimation
radius $R_{\rm dust}\sim 2868 ~~\rm (lt-days)$, which is much larger
than the radius indicated by the \feii emission lag time.

\citet{Kaspi00} measured the \hb flux between 6120\AA\ and 6410 \AA\
in the observational frame (also in Peterson et al. 2004). Their \hb
fluxes include \feii contribution. \citet{Kaspi00} gave \hb flux as
$(18.88\pm0.99)\times 10^{-14} \ergs \rm cm^{-2}$. Removing \feii
contamination in this region, the \hb flux is $(14.22\pm1.01)\times
10^{-14} \ergs \rm cm^{-2}$. With our new light curve for \hb, we
cannot determine the time lag for \hb emission line, and we can
determine the time lag for \feii line. It is possible due to: (1)\hb
line has a asymmetric profile (Figure 2; narrow and broad
components), suggesting that \hb is coming from the region with a
very broad size and \feii is coming from the region with a narrow
size. Therefore, we can detect \feii time lag. The \hb flux
in\citet{Kaspi00} and \citet{Peterson04} includes \feii
contribution. (2)\citet{Peterson04} found two peaks in its CCCD (see
their Figure 15) and suggested that the peak at zero is due to
correlated error. The spectral S/Ns are not high and the time
sampling in the light curves is not good. More data and higher S/N
spectra are needed in the future.

With respect to the \hb time lag of 252 days suggested by
\citet{Peterson04}, smaller estimated \hb time lag of 148 days,
which leads to the smaller black hole mass estimation in the
logarithm, is decreased by 0.23 dex. However, considering the
uncertainties of time lag and the mass calculation, our results are
consistent with that from \citet{Peterson04}.

\subsection{The relation between the spectral index and \lv}
   \begin{figure}
   \centering
   \includegraphics[height=7cm,angle=0]{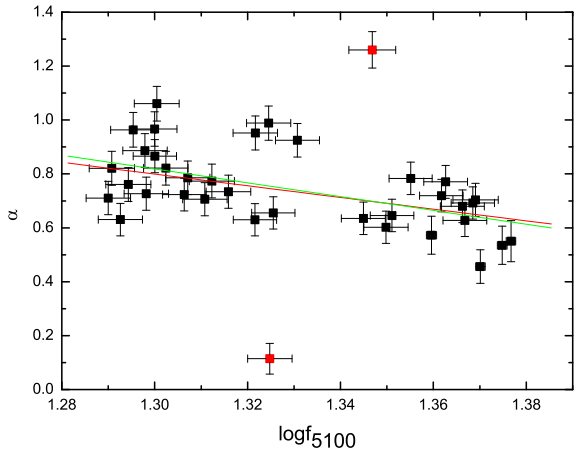}
   \caption{Spectral slope, $\alpha$, vs. \lv, $f_{\nu} \propto \nu ^{-\alpha}$.
   Considering the error of $\alpha$, the red line is the best linear fit.
   The green line is the best linear fit excluding two discrete red points. }
   \end{figure}
In our previous work \citet{Pu06}, we investigated the relation
between the spectral index and \lv and found almost all (15/17) PG
QSOs showed an anti-correlation between them, except PG 1700+518 and
PG 1229+204. For PG 1229+204, it is due to the system difference in
two telescopes. For PG 1700+518, we suggested that it is maybe due
to the \feii contribution \citep{Pu06}. Here, we give this relation
for PG 1700+518 when \feii contribution is carefully removed. We
find that there is a strong anti-correlation between them (see
Figure 6). The spearman coefficient $R$ is -0.33, with a probability
of $p_{\rm null} < 0.05$ for rejecting the null hypothesis of no
correlation. If two discrete red points are excluded, $R$ is 0.55,
and $p_{\rm null} < 5.8\times 10^{-4}$. Therefore, after considering
\feii contribution, PG 1700+518 shares the same characteristic on
spectral slope variability as other 15 PG QSOs in our previous paper
 \citep{Pu06}, i.e., harder spectrum during brighter phase
(Hubeny et al. 2000).

\section{Conclusion}
With the spectral decomposition of 39 spectra of PG 1700+518 with
the strong \feii emission, we investigate the \feii variability and
its time lag. The main conclusions can be summarized as follows: (1)
we give light curves of \lv, \feii, \hbb, \hbn, and \hb$^{n+b}$, as
well as the mean and rms spectra for PG 1700+518. With the
normalized variability measure, $\sigma_N$, we find that all
components are variable. (2) With the code of Peterson et al.
(2004), we find that \feii time lag in PG1700+518 is
$209^{+100}_{-147}$ days, and \hb time lag cannot be determined. (3)
Considering the uncertainties of time lags, the expected \hb time
lag from the empirical luminosity--size relation is 221.6 lt-days,
consistent with our measured \feii time lag. If we take FWHM/time
lag uncertainties into consideration, \feii emission region is
located near the \hb emission region, not conclusively located
outside of the \hb emission region. (4) Assuming that \feii and \hb
emission regions follow the virial relation between the time lag and
the FWHM for the \hb and \feii emission lines, we can derive that
the \hb time lag is $148^{+72}_{-104}$ days. With respect to the \hb
time lag of 252 days suggested by Peterson et al. (2004), smaller
\hb time lag, which leads to the black hole mass estimation in the
logarithm, is decreased by 0.23 dex. (5) After considering \feii
contribution, PG 1700+518 shares the same characteristic on spectral
slope variability to other 15 PG QSOs in our previous work
\citep{Pu06}, i.e., harder spectrum during brighter phase.

\section*{ACKNOWLEDGMENTS}
We are very grateful to B. M. Peterson for his code to determine the
time delay and its error. We thank discussions among people in IHEP
AGN group. We thank an anonymous referee for suggestions that led to
improvements in this paper. This work has been supported by the NSFC
(grants 10873010, 10733010 and 10821061), CAS-KJCX2-YW-T03, and the
National Basic Research Programme of China - the 973 Programme
(grant 2009CB824800).

\bibliography{}

\end{document}